\documentclass[twocolumn,prl,aps,superscriptaddress,showpacs]{revtex4}
\usepackage{graphicx}
\newcommand{\sss}{\scriptscriptstyle}
\begin{document}                                                
\begin{titlepage}
\begin{flushright}BNL-HET-02/27
\end{flushright}
\begin{flushright}FSU-HEP-2002/1115
\end{flushright} 
\begin{flushright}UB-HET-02/09
\end{flushright}
\begin{flushright}hep-ph/0211438
\end{flushright}
\vspace{1truecm}
\begin{center}
{\large\bf
Associated Top Quark-Higgs Boson Production at the LHC 
}
\\
\vspace{0.5in}
{\bf S.~Dawson}\\
{\it Department of Physics, Brookhaven National Laboratory,\\
Upton, NY 11973-5000, USA}
\\ 
\vspace{.25in}
{\bf L.~H.~Orr}\\
{\it Department of Physics $\&$ Astronomy, University of Rochester, \\
Rochester, NY 14627-0171, USA}
\\
\vspace{.25in}
{\bf L.~Reina}\\
{\it Department of Physics, Florida State University,\\ 
Tallahassee, FL 32306-4350, USA}
\\
\vspace{.25in}
{\bf D.~Wackeroth}\\
{\it  Department of Physics, State University of New York at Buffalo, \\
Buffalo, NY 14260-1500, USA} 
\vspace{2in}  

{\bf Abstract}
\end{center}
  We compute the ${\cal O}(\alpha_s^3)$ inclusive cross section for
  the process $pp\to t\bar{t}h$ in the Standard Model, at
  $\sqrt{s_{\sss H}}\!=\!14$~TeV. The next-to-leading order
  corrections drastically reduce the renormalization and factorization
  scale dependence of the Born cross section and increase the total
  cross section for renormalization and factorization scales larger
  than $m_t$.  These corrections have important implications for
  models of new physics involving the top quark.
\end{titlepage}
\vspace*{5cm}
\newpage

\setcounter{page}{0}
\title{Associated Top Quark-Higgs Boson Production at the LHC}
\author{S.~Dawson}
\affiliation{Department of Physics, Brookhaven National Laboratory,
Upton, NY 11973-5000, USA}
\author{L.~H.~Orr}
\affiliation{Department of Physics $\&$ Astronomy, University of Rochester,
Rochester, NY 14627-0171, USA}
\author{L.~Reina}
\affiliation{Department of Physics, Florida State University, 
Tallahassee, FL 32306-4350, USA}
\author{D.~Wackeroth}
\affiliation{Department of Physics, State University of New York at Buffalo,
Buffalo, NY 14260-1500, USA}
\date{\today}
\begin{abstract} 
  We compute the ${\cal O}(\alpha_s^3)$ inclusive cross section for
  the process $pp\to t\bar{t}h$ in the Standard Model, at
  $\sqrt{s_{\sss H}}\!=\!14$~TeV. The next-to-leading order
  corrections drastically reduce the renormalization and factorization
  scale dependence of the Born cross section and increase the total
  cross section for renormalization and factorization scales larger
  than $m_t$.  These corrections have important implications for
  models of new physics involving the top quark.
\end{abstract}
\maketitle

\paragraph{{\bf 1.}}
One of the major goals of the LHC is uncovering the mechanisms of
electroweak symmetry breaking and the generation of fermion masses.
In the Standard Model of particle physics, the masses of gauge bosons
and fermions are generated by a single scalar field.  After
spontaneous symmetry breaking, a neutral CP-even Higgs boson, $h$,
remains as a physical particle. The fermion masses then arise through
couplings to the Higgs boson. In the Standard Model, this coupling is
directly proportional to the fermion mass.  Since the top quark is the
most massive quark, its coupling to the Higgs boson is particularly
sensitive to the underlying physics.
  
The associated production of a Higgs boson with a $t\bar{t}$ pair at
the LHC, $pp\to t\bar{t}h$, will play a very important role in the
115~GeV$\le\!M_h\!\le$140~GeV Higgs boson mass range, both for
discovery and for precision measurements of the Higgs boson couplings.
This process will provide a direct measurement of the top-quark Yukawa
coupling and will be instrumental in determining ratios of Higgs boson
couplings in a model independent
way~\cite{Belyaev:2002ua,Zeppenfeld:2002ng}. Such measurements could
help to distinguish a SM Higgs boson from more complex Higgs sectors,
e.g., as predicted by supersymmetry, and shed light on the details of
the generation of fermion masses.
  
In order to interpret the evidence for $t\bar{t}h$ production and the
measurement of the $t\bar{t}h$ coupling as a verification of the
Standard Model or as a signal for new physics, it is necessary to have
a precise prediction for the cross section.  QCD corrections are
expected to be important and are crucial in order to reduce the
dependence of the cross section on the arbitrary renormalization and
factorization scales.  In this letter, we present the
next-to-leading-order (NLO) QCD corrections to the total cross section
for $pp\to t\bar{t}h$ at the LHC.  Our results are in very good
agreement with those of Ref.~\cite{Beenakker:2001rj} within the
statistical errors.  Results for the Fermilab Tevatron have been
presented elsewhere \cite{Beenakker:2001rj,Reina:2001sf,Reina:2001bc}.

\paragraph{{\bf 2.}}
The inclusive total cross section for $pp\to t\bar{t}h$ at ${\cal
  O}(\alpha_s^3)$ can be written as:
\begin{eqnarray}
\label{eq:sigma_nlo}
\lefteqn{\sigma_{\sss NLO}(p\,p 
\to t\bar{t}h) =\sum_{ij} \frac{1}{1+\delta_{ij} } \int dx_1 dx_2}  \\
&& \hspace*{-0.1cm}\cdot[{\cal F}_i^p(x_1,\mu) {\cal F}_j^{p}(x_2,\mu)
{\hat \sigma}^{ij}_{\sss NLO}(x_1,x_2,\mu) + (1\leftrightarrow 2)]
\,\,\,,\nonumber
\end{eqnarray}
where $ {\cal F}_i^{p}$ is the NLO parton distribution function for
parton $i$ in a proton, defined at a factorization scale
$\mu_f\!=\!\mu$, and ${\hat \sigma}^{ij}_{\sss NLO}$ is the ${\cal
  O}(\alpha_s^3)$ parton level total cross section for incoming
partons $i$ and $j$, made of the channels $q\bar{q},\, gg\to
t\bar{t}h$, and $(q,\bar q)g\to t\bar{t}h(q,\bar q)$,
and renormalized at the scale $\mu_r$ which we also take to be
$\mu_r\!=\!\mu$.  At the LHC, the dominant contribution is from the
gluon-gluon initial state, although the other contributions cannot be
neglected and are included in this calculation.

The NLO parton-level total cross section, ${\hat \sigma}^{ij}_{\sss
  NLO}$, consists of the ${\cal O}(\alpha_s^2)$ Born cross section,
${\hat \sigma}^{ij}_{\sss LO}$, and the ${\cal O}(\alpha_s)$
corrections to the Born cross section, $\delta {\hat
  \sigma}^{ij}_{\sss NLO}$, including the effects of mass
factorization. $\delta {\hat\sigma}^{ij}_{\sss NLO}$ contains virtual
and real corrections to the parton-level $t\bar{t}h$ production
processes, $q\bar{q}\to t\bar{t}h$ and $gg\to t\bar{t}h$, and the
tree-level $(q,\bar{q})g$ initiated processes, $(q,\bar{q})g\to
t\bar{t}h(q,\bar{q})$, which are of the same order in $\alpha_s$.  It
can be written as the sum of two terms:
\begin{eqnarray}
\label{eq:dsigma_nlo}
&& \delta\hat{\sigma}^{ij}_{\sss NLO} = {\hat\sigma}_{virt}^{ij}
+{\hat\sigma}_{real}^{ij}=\\
&&\int d(PS_3) M(ij\to t\bar{t}h)
+\int d(PS_4) M({ij}\to t\bar{t}h+k)\,\,\,,\nonumber
\end{eqnarray}
where $M(ij\to t\bar{t}h)$ and $M(ij\to t\bar{t}h+k)$ (for
$k\!=\!g,q,\bar{q}$) are respectively the matrix elements squared for
the ${\cal O}(\alpha_s^3)$ $2\to 3$ and $2\to 4$ scattering processes
averaged over the initial degrees of freedom and summed over the final
ones, while $d(PS_3)$ and $d(PS_4)$ denote the integration over the
corresponding three/four particle phase space.  The $(q,\bar{q})g$
initial state contributes only to $\hat{\sigma}^{ij}_{real}$.

The main challenges in the calculation come from the presence in the
virtual corrections of pentagon diagrams with several massive external
and internal particles, and from the computation of the real part in
the presence of infrared singularities.

\paragraph{{\bf 3.}}  
The ${\cal O}(\alpha_s)$ virtual corrections to the tree level $ ij\to
t\bar{t}h$ ($ij=q\bar{q},~gg$) processes consist of self-energy,
vertex, box, and pentagon diagrams.  The calculation of the virtual
corrections to the $q\bar{q}$ initial state is described in
Ref.~\cite{Reina:2001bc}.  The basic method is to reduce each diagram
to a sum of scalar integrals of the form,
\begin{equation}
\label{eq:scalar_int}
\int {d^dk\over (2\pi)^d}
\prod_{i=0}^{n\le4} {1\over [(k+p_i)^2-m_i^2]}\,\,\, ; \; p_0=0 \;,
\end{equation}
that may contain both ultraviolet (UV) and infrared (IR)
singularities.  The finite scalar integrals are evaluated by using the
method described in Ref.~\cite{Denner:1993kt} and cross checked with
the numerical package FF~\cite{vanOldenborgh:1990wn}.  The scalar
integrals that exhibit UV and/or IR divergences are calculated
analytically. Both the UV and IR divergences are extracted by using
dimensional regularization in $d\!=\!4-2\epsilon$ dimensions. The UV
divergences are then removed by introducing a suitable set of
counterterms, as described in detail in Ref.~\cite{Reina:2001bc}. The
IR divergences are cancelled by the analogous singularities in the
soft and collinear part of the real gluon emission cross section.

The most difficult integrals arise from the IR divergent pentagon
diagrams with several massive particles.  In
Refs.~\cite{Reina:2001sf,Reina:2001bc,Dawson:2002kn} we calculated the
pentagon scalar integrals as linear combinations of scalar box
integrals using the method of Ref.~\cite{Bern:1993em,Bern:1994kr}.
For the $gg$ initiated process we also used the method of
Ref.~\cite{Denner:1993kt} and found perfect agreement between the
results of the two methods.  The virtual corrections to the $gg$
initiated process have an additional complication with respect to the
$q\bar{q}$ case because of the presence of pentagon tensor integrals
with rank higher than one.  Pentagon tensor integrals can give rise to
numerical instabilities due to the dependence on inverse powers of the
Gram determinant (GD):
\begin{eqnarray}
\label{eq:gram_det}
\lefteqn{{\rm GD}(p_1+p_2\to \sum_{i=3}^5 p_i)
=-\frac{[s-(2 m_t+M_h)^2]}{64}
 \times} \nonumber \\
&& [M_h^4  + (s - s_{45})^2  - 2 M_h^2  (s + s_{45})] s s_{45}
\times \nonumber \\
&&\sin^2\theta_{45} \sin^2\phi_{45} \sin^2\theta,
\end{eqnarray}
where the 3-particle phase space has been expressed in terms of a
time-like invariant $s_{45}=(p_4+p_5)^2$, polar angles,
$\theta_{45},\theta$ and azimuthal angles, $\phi_{45},\phi$, and
$s\!=\!x_1x_2s_{\sss H}$ is the partonic center-of-mass energy
squared.  As can be seen in Eq.~(\ref{eq:gram_det}), the Gram
determinant vanishes when two momenta become degenerate, i.e. at the
boundaries of phase space.  These are spurious divergences, which
cause serious numerical difficulties.  We use two methods to overcome
this problem and find mutual agreement within the statistical
uncertainty of the phase space integration:
\begin{itemize}
\item Impose kinematic cuts to avoid the phase space regions where the
  Gram determinant vanishes.  Then apply an extrapolation procedure
  from the numerically safe to the numerically unsafe region.
\item Eliminate all pentagon tensor integrals by cancelling terms in
  the numerator against the propagators wherever possible, after
  interfering the pentagon amplitude with the Born matrix element. The
  resulting expressions are very large, but numerically stable.
\end{itemize}

\paragraph{{\bf 4.}} 
The ${\cal O}(\alpha_s)$ corrections to the Born cross sections for
the $q\bar{q}$ and $gg$ initial states due to real gluon emission, as
well as the $(q,\bar{q})g$ initiated processes, have been computed
using both a two cutoff~\cite{Harris:2001sx} and a single
cutoff~\cite{Giele:1992vf,Giele:1993dj,Keller:1998tf} implementation
of the phase space slicing (PSS) algorithm.  In both PSS methods, the
real contribution to the NLO rate is computed analytically below the
cutoff(s) and numerically above the cutoff(s) and the final result is
independent of these arbitrary parameters.  When studying the cutoff
dependence of $\sigma_{\sss NLO}$, it is crucial to choose the
cutoff(s) small enough to justify the analytical calculations of the
IR divergent contributions to ${\hat\sigma}_{real}^{ij}$, but not so
small as to cause numerical instabilities.  Finding agreement between
the two PSS approaches is therefore a strong check of the accuracy of
the calculation.

In the two cutoff PSS algorithm, the contributions of $q\bar q,gg\to
t\bar{t}h+g$ to $\hat{\sigma}_{real}^{ij}$ are first divided into a
soft and a hard part,
\begin{equation}
\label{eq:soft_hard}
\hat\sigma_{real}^{ij}=\hat\sigma_{soft}^{ij}+\hat\sigma_{hard}^{ij}
\,\,\,,
\end{equation}
where \emph{soft} and \emph{hard} refer to the energy of the final
state radiated gluon.  This division into hard and soft contributions
depends on a soft cutoff, $\delta_s$, such that the energy of the
radiated gluon in the partonic center-of-mass frame is considered soft
if $E_g\le\delta_s\frac{\sqrt{s}}{2}$.  The eikonal approximation to
the soft matrix elements can be taken and the integral over the soft
degrees of freedom performed analytically.  Since the $(q,\bar{q})g$
initiated process does not develop soft singularities, it only
contributes to $\hat{\sigma}^{qg}_{hard}$ and no soft cutoff is
applied.

The hard contribution to $ij\to t\bar{t}h+k$ ($k=g,q,\bar q$) is
further divided into a hard/collinear part,
$\hat\sigma_{hard/coll}^{ij}$, and a hard/non collinear part,
$\hat\sigma_{hard/non-coll}^{ij}$.  For $q\bar q$ and $gg$ initiated
processes the hard/collinear region is defined as the region where the
energy of the final state gluon is $E_g>\delta_s\frac{\sqrt{s}}{2}$
and the gluon is radiated from the initial massless parton at an angle
$\theta_{ig}$, in the $ij$ center-of-mass frame, such that
$(1-\cos\theta_{gi})\le\delta_c$, for an arbitrary small collinear
cutoff $\delta_c$.  The matrix element squared in the hard/collinear
limit is calculated using the leading pole approximation, resulting
into the convolution of the unregulated Altarelli-Parisi splitting
functions $P_{q\bar{q}}$ or $P_{gg}$ with the corresponding tree level
matrix elements squared. In the same way, the matrix elements squared
for the $(q,\bar q)g$ initiated processes are described, in the
collinear region $(1-\cos\theta_{(q,\bar q)i})\!\le\!\delta_c$, as
convolutions of the unregulated Altarelli-Parisi splitting functions
$P_{gq}$ and $P_{qg}$ with the $q\bar q$ and $gg$ tree-level matrix
elements, respectively.  The integration over the angular degrees of
freedom can then be performed analytically.

The hard gluon emission from the final massive quarks never belongs to
the hard/collinear region.  The contribution from the hard/non
collinear region is finite and is computed numerically, using standard
Monte Carlo integration techniques.

\begin{figure}[htb]
\begin{center}
\includegraphics[width=16pc]{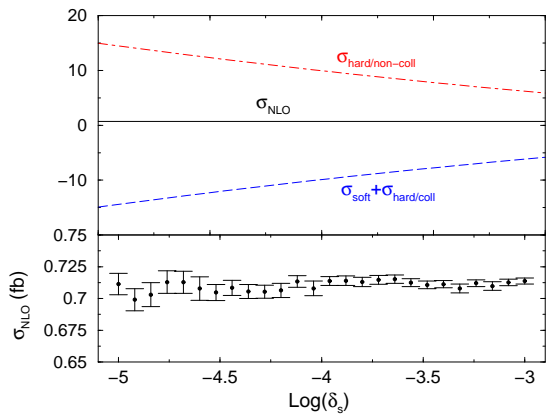} 
\vspace*{-0.5truecm}
\caption[]{
  $\sigma_{\sss NLO}(pp\to t\bar{t}h)$ calculated in the two cutoff
  PSS approach when varying the soft cutoff $\delta_s$ in the range
  $10^{-5}$-$10^{-3}$, with the collinear cutoff set to
  $\delta_c\!=\!\delta_s/100$, at $\sqrt{s_{\sss H}}\!=\!14$~TeV, for
  $M_h\!=\!120$~GeV and $\mu\!=\!m_t$.  The upper plot shows the
  cancellation of the $\delta_s,\delta_c$ dependence between
  $\sigma_{soft}+\sigma_{hard/coll}$ and $\sigma_{hard/non-coll}$. The
  lower plot shows the dependence of $\sigma_{\sss NLO}$ on
  $\delta_s,\delta_c$, with the corresponding statistical errors.}
\label{fg:deltas}
\end{center}
\end{figure}

Both $\hat\sigma_{soft}^{ij}$ and $\hat\sigma_{hard}^{ij}$ depend on
the arbitrary cutoff $\delta_s$, and ${\hat \sigma}_{hard/coll}^{ij}$
and ${\hat\sigma}_{hard/non-coll}^{ij}$ also depend on $\delta_c$.
However, the real hadronic cross section, $\sigma_{real}$, after mass
factorization, is cutoff independent. The cutoff independence of the
NLO cross section, $\sigma_{\sss NLO}$, is shown in
Fig.~\ref{fg:deltas}, where we assume $\delta_c\!=\!\delta_s/100$ and
we let $\delta_s$ vary between $10^{-5}$ and $10^{-3}$. The hadronic
cross sections $\sigma_{soft}+\sigma_{hard/coll}$ and
$\sigma_{hard/non-coll}$ include the contributions of all three $t\bar
t h$ production channels, $q\bar q,gg\to t\bar thg$ and $(q,\bar q)g
\to t\bar th(q,\bar q)$.  For $\delta_s$ in the range
$10^{-5}\!-\!10^{-3}$ and for $\delta_c\!=\!\delta_s/100$, a clear
plateau is reached and the result is independent of $\delta_s$ and
$\delta_c$.
\begin{figure}[htb]
\begin{center}
\includegraphics[width=16pc]{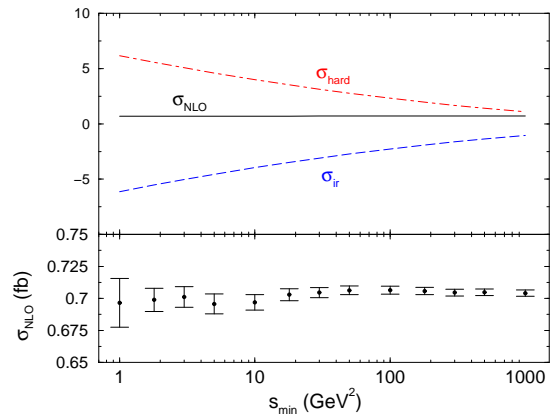}
\vspace*{-0.5truecm}
\caption[ ]{Dependence of $\sigma_{\sss NLO}(pp
  \to t\bar{t}h)$ on the arbitrary cutoff $s_{min}$ of the one
  cutoff PSS method at $\sqrt{s_{\sss H}}\!=\!14$~TeV, for
  $M_h\!=\!120$~GeV and $\mu\!=\!m_t$.  The upper plot shows the
  cancellation of the $s_{min}$ dependence between
  $\sigma_{ir}\!=\!\sigma_{ir}^c+ \sigma_{crossing}$, and
  $\sigma_{hard}$.  The lower plot shows the dependence of
  $\sigma_{\sss NLO}$ on $s_{min}$, with the corresponding statistical
  errors.}
\label{fg:smin_lhc}
\end{center}
\end{figure}

An alternative method of isolating both soft and collinear
singularities is to divide the phase space of the final state partons
into two regions, according to whether all partons can be resolved or
not.
  
The single cutoff PSS technique defines the IR divergent region as
that where the final state parton $k$ emitted from parton $i$ is not
resolved and
\begin{equation}
s_{ik} = 2 p_i\cdot p_k < s_{min}
\end{equation}
for an arbitrarily small value of the cutoff $s_{min}$.  The partonic
real cross section is then written as
\begin{equation}
{\hat \sigma}_{real}^{ij}={\hat \sigma}_{ir}^{ij}+{\hat \sigma}_{hard}^{ij}
={\hat \sigma}_{ir}^{c,ij}+{\hat \sigma}_{crossing}^{ij} +
{\hat \sigma}_{hard}^{ij},
\end{equation}
where ${\hat\sigma}_{ir}^{ij}$ includes both soft and collinear
singularities and ${\hat\sigma}_{hard}^{ij}$ is finite.  The IR
divergent contribution (${\hat\sigma}_{ir}^{c,ij}$) is computed from
the crossed process, $h\rightarrow ijt {\overline t} +k$, where $ij$
denote the initial state partons and $k\!=\!g,q,\bar{q}$. This permits
a straightforward decomposition of the amplitude in terms of color
ordered amplitudes.  Using the factorization properties of both the
color ordered amplitudes and the gluon/quark phase space in the
soft/collinear limits, the IR divergent contribution can be extracted
analytically.  The factorization of soft and collinear singularities
for color ordered amplitudes has been discussed in the literature
mainly for the leading color terms, ${\cal
  O}(N_c)$~\cite{Giele:1992vf,Giele:1993dj,Keller:1998tf}. In our
case, however, the inclusion of the sub-leading terms in $1/N_c$ is
crucial. For the $q\bar{q}$ initial state, these terms were first
calculated in Ref.~\cite{Reina:2001bc}.

An additional contribution to ${\hat\sigma}_{ir}^{ij}$
(${\hat\sigma}_{crossing}^{ij}$) arises when crossing partons $i$ and
$j$ back to the initial state, due to the mismatch between the
collinear gluon radiation from initial and final state partons.
${\hat\sigma}_{crossing}^{ij}$ contains collinear divergences which
are cancelled by the parton distribution counterterms when the parton
cross section is convoluted with the PDF's. In Fig.~\ref{fg:smin_lhc},
the independence of $\sigma_{\sss NLO}$ from the cutoff $s_{min}$ is
demonstrated. Again, the hadronic cross sections
$\sigma_{ir}^c+\sigma_{crossing}$ and $\sigma_{hard}$ include the
contributions from all initial states, $q\bar{q},gg$ and
$(q,\bar{q})g$. Together with Ref.~\cite{Reina:2001bc}, this is the
first application of the single cutoff PSS approach the calculation of
a cross section involving more than one massive particle in the final
state.
  
The numerical results of both PSS methods agree within the statistical
errors and within the systematic errors of the applied soft and
collinear approximations.  In Ref.~\cite{Beenakker:2001rj}, the dipole
subtraction formalism has been used to extract the IR singularities of
the real part.  The agreement between these three very different
treatments of the real IR singularities represents a powerful check of
the corresponding NLO calculations.

\paragraph{{\bf 5.}}
Our numerical results are found using CTEQ4M parton distribution
functions for the calculation of the NLO cross section, and CTEQ4L
parton distribution functions for the calculation of the lowest order
cross section \cite{Lai:1997mg}.  The NLO (LO) cross section is
evaluated using the $2$ ($1$)-loop evolution of $\alpha_s(\mu)$. The
top quark mass is taken to be $m_t\!=\!174$~GeV and $\alpha_s^{\sss
  NLO}(M_Z)\!=\!0.116$.
 
\begin{figure}[hbt]
\begin{center}
\includegraphics[width=16pc]{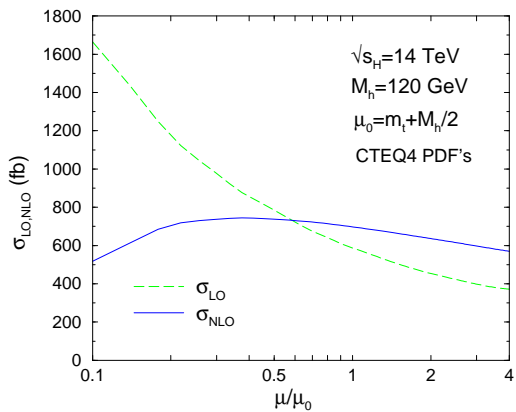}
\vspace*{-0.5truecm}
\caption[ ]{$\sigma_{\sss LO,NLO}(pp\to
  t\bar{t}h)$ as functions of the renormalization/factorization scale
  $\mu$, at $\sqrt{s_{\sss H}}\!=\!14$~TeV, for $M_h\!=\!120$~GeV. }
\label{fg:mudep}
\end{center}
\end{figure}

In Fig.~\ref{fg:mudep} we show, for $M_h\!=\!120$~GeV, the dependence
of $\sigma_{\sss LO}$ and $\sigma_{\sss NLO}$ on the arbitrary
renormalization/factorization scale $\mu\!=\!\mu_r\!=\!\mu_f$. For
scales $\mu$ larger than $m_t$ the NLO result is significantly larger
than the lowest order result.
\begin{figure}[htb]
\begin{center}
\includegraphics[width=16pc]{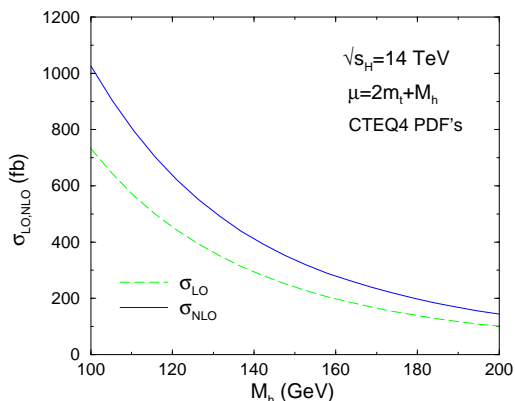}
\vspace*{-0.5truecm}
\caption[ ]{$\sigma_{\sss NLO}$ and $\sigma_{\sss LO}$  for
  $pp\to t\bar{t}h$ as functions of $M_h$, at
  $\sqrt{s_{H}}\!=\!14$~TeV, for $\mu\!=\!2m_t+M_h$.}
\label{fg:mhdep}
\end{center}
\end{figure}

Fig.~\ref{fg:mhdep} shows both the LO and the NLO total cross section
for $p p\rightarrow t\bar th$ at $\sqrt{s_{ H}}\!=\!14$~TeV, as
functions of $M_h$, for a representative value of the
renormalization/factorization scale, $\mu\!=\!2m_t+M_h$. We estimate
the remaining theoretical error, due to the residual $\mu$-dependence,
to the parton distribution functions, and to the experimental error on
$m_t$, to be of the order of 15-20\%. In comparison, the statistical
error on the numerical integration is negligible, due to the high
statistics used in evaluating the total cross section.

\paragraph{{\bf 6.}}
The NLO QCD corrections to the Standard Model process $pp\to
t\bar{t}h$, at $\sqrt{s_{\sss H}}\!=\!14$~TeV increase the LO cross
section by a factor of $1.2\!-\!1.4$ for renormalization/factorization
scales in the range $m_t+M_h/2 \le \mu \le 4 m_t+2 M_h$ and Higgs boson masses 
in the range considered in this paper.
The NLO result shows a drastically reduced scale dependence as
compared to the Born result and leads to increased confidence in
predictions based on these results. The techniques developed in this
calculation can now be applied to the study of the associated
$b\bar{b}h$ production at both the LHC and the Tevatron.

\begin{acknowledgments}
  We thank U.~Baur, Z.~Bern, and F.~Paige for valuable discussions and
  encouragement. We are grateful to the Authors of
  Ref.~\cite{Beenakker:2001rj} for a detailed comparison of the
  results.  The work of S.D. (L.H.O., L.R.) is supported in part by
  the U.S.  Department of Energy under grant DE-AC02-76CH00016
  (DE-FG-02-91ER40685, DE-FG02-97ER41022).
\end{acknowledgments}

\bibliography{tth_lhc_hep}

\begin{thebibliography}{15}
\expandafter\ifx\csname natexlab\endcsname\relax\def\natexlab#1{#1}\fi
\expandafter\ifx\csname bibnamefont\endcsname\relax
  \def\bibnamefont#1{#1}\fi
\expandafter\ifx\csname bibfnamefont\endcsname\relax
  \def\bibfnamefont#1{#1}\fi
\expandafter\ifx\csname citenamefont\endcsname\relax
  \def\citenamefont#1{#1}\fi
\expandafter\ifx\csname url\endcsname\relax
  \def\url#1{\texttt{#1}}\fi
\expandafter\ifx\csname urlprefix\endcsname\relax\def\urlprefix{URL }\fi
\providecommand{\bibinfo}[2]{#2}
\providecommand{\eprint}[2][]{\url{#2}}

\bibitem[{\citenamefont{Belyaev and Reina}(2002)}]{Belyaev:2002ua}
\bibinfo{author}{\bibfnamefont{A.}~\bibnamefont{Belyaev}} \bibnamefont{and}
  \bibinfo{author}{\bibfnamefont{L.}~\bibnamefont{Reina}},
  \bibinfo{journal}{JHEP} \textbf{\bibinfo{volume}{08}}, \bibinfo{pages}{041}
  (\bibinfo{year}{2002}), \eprint[http://arXiv.org/abs]{hep-ph/0205270}.

\bibitem[{\citenamefont{Zeppenfeld}(2002)}]{Zeppenfeld:2002ng}
\bibinfo{author}{\bibfnamefont{D.}~\bibnamefont{Zeppenfeld}}
  (\bibinfo{year}{2002}), \eprint[http://arXiv.org/abs]{hep-ph/0203123}.

\bibitem[{\citenamefont{Beenakker et~al.}(2001)\citenamefont{Beenakker,
  Dittmaier, Kr{\"a}mer, Pl{\"u}mper, Spira, and Zerwas}}]{Beenakker:2001rj}
\bibinfo{author}{\bibfnamefont{W.}~\bibnamefont{Beenakker}},
  \bibinfo{author}{\bibfnamefont{S.}~\bibnamefont{Dittmaier}},
  \bibinfo{author}{\bibfnamefont{M.}~\bibnamefont{Kr{\"a}mer}},
  \bibinfo{author}{\bibfnamefont{B.}~\bibnamefont{Pl{\"u}mper}},
  \bibinfo{author}{\bibfnamefont{M.}~\bibnamefont{Spira}}, \bibnamefont{and}
  \bibinfo{author}{\bibfnamefont{P.}~\bibnamefont{Zerwas}},
  \bibinfo{journal}{Phys. Rev. Lett.} \textbf{\bibinfo{volume}{87}},
  \bibinfo{pages}{201805} (\bibinfo{year}{2001}),
  \eprint[http://arXiv.org/abs]{hep-ph/0107081};
  \bibinfo{note}{hep-ph/0211352}.

\bibitem[{\citenamefont{Reina and Dawson}(2001)}]{Reina:2001sf}
\bibinfo{author}{\bibfnamefont{L.}~\bibnamefont{Reina}} \bibnamefont{and}
  \bibinfo{author}{\bibfnamefont{S.}~\bibnamefont{Dawson}},
  \bibinfo{journal}{Phys. Rev. Lett.} \textbf{\bibinfo{volume}{87}},
  \bibinfo{pages}{201804} (\bibinfo{year}{2001}),
  \eprint[http://arXiv.org/abs]{hep-ph/0107101}.

\bibitem[{\citenamefont{Reina et~al.}(2002)\citenamefont{Reina, Dawson, and
  Wackeroth}}]{Reina:2001bc}
\bibinfo{author}{\bibfnamefont{L.}~\bibnamefont{Reina}},
  \bibinfo{author}{\bibfnamefont{S.}~\bibnamefont{Dawson}}, \bibnamefont{and}
  \bibinfo{author}{\bibfnamefont{D.}~\bibnamefont{Wackeroth}},
  \bibinfo{journal}{Phys. Rev.} \textbf{\bibinfo{volume}{D65}},
  \bibinfo{pages}{053017} (\bibinfo{year}{2002}),
  \eprint[http://arXiv.org/abs]{hep-ph/0109066}.

\bibitem[{\citenamefont{Denner}(1993)}]{Denner:1993kt}
\bibinfo{author}{\bibfnamefont{A.}~\bibnamefont{Denner}},
  \bibinfo{journal}{Fortschr. Phys.} \textbf{\bibinfo{volume}{41}},
  \bibinfo{pages}{307} (\bibinfo{year}{1993}).

\bibitem[{\citenamefont{van Oldenborgh and
  Vermaseren}(1990)}]{vanOldenborgh:1990wn}
\bibinfo{author}{\bibfnamefont{G.~J.} \bibnamefont{van Oldenborgh}}
  \bibnamefont{and} \bibinfo{author}{\bibfnamefont{J.~A.~M.}
  \bibnamefont{Vermaseren}}, \bibinfo{journal}{Z. Phys.}
  \textbf{\bibinfo{volume}{C46}}, \bibinfo{pages}{425} (\bibinfo{year}{1990}).

\bibitem[{\citenamefont{Dawson et~al.}(2002)\citenamefont{Dawson, Orr, Reina,
  and Wackeroth}}]{Dawson:2002kn}
\bibinfo{author}{\bibfnamefont{S.}~\bibnamefont{Dawson}},
  \bibinfo{author}{\bibfnamefont{L.H.}~\bibnamefont{Orr}},
  \bibinfo{author}{\bibfnamefont{L.}~\bibnamefont{Reina}}, \bibnamefont{and}
  \bibinfo{author}{\bibfnamefont{D.}~\bibnamefont{Wackeroth}}
  (\bibinfo{year}{2002}), \eprint[http://arXiv.org/abs]{hep-ph/0210109}.

\bibitem[{\citenamefont{Bern et~al.}(1993)\citenamefont{Bern, Dixon, and
  Kosower}}]{Bern:1993em}
\bibinfo{author}{\bibfnamefont{Z.}~\bibnamefont{Bern}},
  \bibinfo{author}{\bibfnamefont{L.~J.} \bibnamefont{Dixon}}, \bibnamefont{and}
  \bibinfo{author}{\bibfnamefont{D.~A.} \bibnamefont{Kosower}},
  \bibinfo{journal}{Phys. Lett.} \textbf{\bibinfo{volume}{B302}},
  \bibinfo{pages}{299} (\bibinfo{year}{1993}), \bibinfo{note}{erratum-ibid.
  {\bf B318}, 649 (1993)}, \eprint[http://arXiv.org/abs]{hep-ph/9212308}.

\bibitem[{\citenamefont{Bern et~al.}(1994)\citenamefont{Bern, Dixon, and
  Kosower}}]{Bern:1994kr}
\bibinfo{author}{\bibfnamefont{Z.}~\bibnamefont{Bern}},
  \bibinfo{author}{\bibfnamefont{L.~J.} \bibnamefont{Dixon}}, \bibnamefont{and}
  \bibinfo{author}{\bibfnamefont{D.~A.} \bibnamefont{Kosower}},
  \bibinfo{journal}{Nucl. Phys.} \textbf{\bibinfo{volume}{B412}},
  \bibinfo{pages}{751} (\bibinfo{year}{1994}),
  \eprint[http://arXiv.org/abs]{hep-ph/9306240}.

\bibitem[{\citenamefont{Harris and Owens}(2002)}]{Harris:2001sx}
\bibinfo{author}{\bibfnamefont{B.~W.} \bibnamefont{Harris}} \bibnamefont{and}
  \bibinfo{author}{\bibfnamefont{J.~F.} \bibnamefont{Owens}},
  \bibinfo{journal}{Phys. Rev.} \textbf{\bibinfo{volume}{D65}},
  \bibinfo{pages}{094032} (\bibinfo{year}{2002}),
  \eprint[http://arXiv.org/abs]{hep-ph/0102128}.

\bibitem[{\citenamefont{Giele and Glover}(1992)}]{Giele:1992vf}
\bibinfo{author}{\bibfnamefont{W.~T.} \bibnamefont{Giele}} \bibnamefont{and}
  \bibinfo{author}{\bibfnamefont{E.~W.~N.} \bibnamefont{Glover}},
  \bibinfo{journal}{Phys. Rev.} \textbf{\bibinfo{volume}{D46}},
  \bibinfo{pages}{1980} (\bibinfo{year}{1992}).

\bibitem[{\citenamefont{Giele et~al.}(1993)\citenamefont{Giele, Glover, and
  Kosower}}]{Giele:1993dj}
\bibinfo{author}{\bibfnamefont{W.~T.} \bibnamefont{Giele}},
  \bibinfo{author}{\bibfnamefont{E.~W.~N.} \bibnamefont{Glover}},
  \bibnamefont{and} \bibinfo{author}{\bibfnamefont{D.~A.}
  \bibnamefont{Kosower}}, \bibinfo{journal}{Nucl. Phys.}
  \textbf{\bibinfo{volume}{B403}}, \bibinfo{pages}{633} (\bibinfo{year}{1993}),
  \eprint[http://arXiv.org/abs]{hep-ph/9302225}.

\bibitem[{\citenamefont{Keller and Laenen}(1999)}]{Keller:1998tf}
\bibinfo{author}{\bibfnamefont{S.}~\bibnamefont{Keller}} \bibnamefont{and}
  \bibinfo{author}{\bibfnamefont{E.}~\bibnamefont{Laenen}},
  \bibinfo{journal}{Phys. Rev.} \textbf{\bibinfo{volume}{D59}},
  \bibinfo{pages}{114004} (\bibinfo{year}{1999}),
  \eprint[http://arXiv.org/abs]{hep-ph/9812415}.

\bibitem[{\citenamefont{Lai et~al.}(1997)}]{Lai:1997mg}
\bibinfo{author}{\bibfnamefont{H.~L.} \bibnamefont{Lai}} \bibnamefont{et~al.},
  \bibinfo{journal}{Phys. Rev.} \textbf{\bibinfo{volume}{D55}},
  \bibinfo{pages}{1280} (\bibinfo{year}{1997}),
  \eprint[http://arXiv.org/abs]{hep-ph/9606399}.

\end{thebibliography}
\end{document}